# Dynamically creating asymmetrical potential wells for nanoparticle manipulation

XIONGGUI TANG,* YANHUA XU

Department of Physics, Key Laboratory of Low Dimensional Quantum Structures and Quantum Control of Ministry of Education, Synergetic Innovation Center for Quantum Effects and Applications, Hunan Normal University, Changsha, 410081, P.R. China
*Corresponding author: tangxg@hunnu.edu.cn

**Optical tweezers, which are powerful tools for trapping and manipulating particles, have been widely used in many areas. However, their potential wells are typically symmetrical, which limit their capability of optical trapping and manipulation. Herein, we report an approach for dynamically creating asymmetrical potential wells, by employing variable phase gradient profiles generated by holograms. Its depth, stiffness and location of potential wells can be dynamically controlled. Moreover, the controllable transport of nanoparticle is experimentally demonstrated, which reveals that it exhibits excellent manipulation performance. Importantly, its motion path, step distance and velocity can be easily manipulated by changing its holograms, which is highly desirable in controllable transport of nanoparticles. This study offers a new tool for optical manipulation, which could largely expand the capability of conventional optical tweezers.**

Optical tweezers, firstly demonstrated by Arthur Ashkin [1], have triggered a promising application in many fields such as quantum physics, biological medicine, chemistry [2-10]. In past two decades, many different techniques have been developed to generate various optical tweezers, including holographic optical tweezers [11-12], plasmonic optical tweezers [13-15], and fiber optical tweezers [16-18]. For these optical tweezers, holographic optical tweezer, coined by Curtis et al in 2002 [19], has become one of the most attractive techniques in optical trapping systems, which is usually generated by using phase-only spatial light modulator (SLM). Since then, different techniques for optical trapping and manipulation have been demonstrated by employing holographic optical tweezers [20-25]. The optical traps are generated by intensity gradient, and their potential wells are usually symmetrical, which largely limit their capability in optical manipulation. Fortunately, the optical force can also originate from transverse phase gradient, which provide a new tool for optical manipulation. Various particle manipulation including optical sorting and self-assemble have been reported, in which the potential wells are also symmetrical [26,27]. Additionally, the intensity and phase profiles are highly coupled in these line traps, which have detrimental effect on optical manipulation. To date, it remains challenging to accurately create asymmetrical potential wells by using transverse phase or intensity gradient.

In contrast, the needs for asymmetrical potential wells in many fields are rapidly increasing. For instance, optical Brownian ratchet is a good example. The motion of particles are manipulated by employing asymmetrical potential wells. Consequently, different schemes have been proposed, to demonstrate optical Brownian ratchets [28-29]. Recently, Shao-Hua Wu et. al. presented novel optical Brownian ratchet for particle manipulation in lab-on-chip environment, by employing near-field on-chip traps on an asymmetrically patterned photonic crystal [30]. Actually, the reported approaches for creating asymmetrical potential wells are realized by using movable spot optical traps or near-field traps. Obviously, There exists several disadvantages, such as low speed, small step distance, large particle volume, the limited motion path or the complicated experimental systems. Accordingly, it is highly desirable to explore asymmetrical potential wells with high accuracy and easy realization.

In this work, we report an approach for generating asymmetrical potential wells, by utilizing phase gradient profiles. Especially, the potential wells can be dynamically controllable by changing holograms. Using dynamic asymmetrical potential wells, we experimentally demonstrate a new controllable transport. It indicates that the manipulation performance is excellent. Apparently, the advantages are remarkable, including flexible design, easy control, high motion speed, large step distance and variable motion path.

In optical trapping system, particles are definitely trapped by optical force, arising from intensity and phase gradient. It can be calculated by [31],

$$\vec{F} = \frac{1}{4}\varepsilon_0\varepsilon\alpha'\nabla I + \frac{1}{2}\varepsilon_0\varepsilon\alpha''I\nabla\varphi, \quad (1)$$

where $\varepsilon_0$ and $\varepsilon$ are dielectric constant and relative dielectric constant, $\alpha'$ and $\alpha''$ are the real and imaginary part of particle's polarizability, $I$ is the optical intensity, and $\varphi$ is the phase profile of optical field. The first term in Eq. (1) is induced by optical intensity gradient profiles, and the second one is generated by phase gradient profiles. If the intensity is constant in the trapping region, the intensity gradient force becomes zero. Therefore, optical force is only determined by phase gradient force. In the following, we can easily obtain its potential energy by,

$$V(x,y) = -k\varphi, \quad (2)$$

where $k = \varepsilon_0 \varepsilon \alpha'' I / 2$. Apparently, there is anti-symmetrical relationship between potential energy profiles and phase profiles. Accordingly, we can design phase profiles for accurately creating the desirable potential energy profiles. In the following, asymmetrical potential energy profiles shown in Fig. 1(a), can be generated by using phase profiles, as demonstrated in Fig. 1(b). Especially, its intensity distribution is uniform along trapping orbit, as depicted in Fig. 1(c). To accurately create the desirable phase profiles, holographic optical trapping system is employed, due to its advantages including easy control, high diffraction efficiency, and great flexibility. In our previous study, an approach for directly calculating holograms has been proposed, which exhibits high accuracy in shaping the intensity and phase profiles of light [32]. Herein, this method is employed to design the holograms, which are used for creating asymmetrical potential wells. Importantly, such complex phase profiles can be accurately obtained in our method. Although the complex phase profiles could have slight negative effect on the quality of intensity distribution, it can be effectively improved, by modifying the holograms.

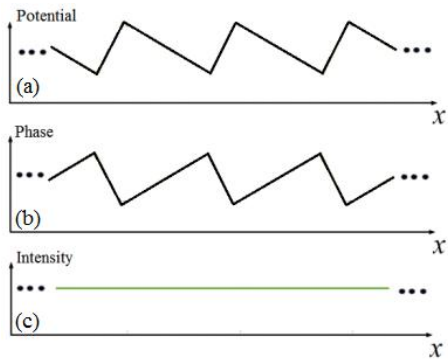

Fig. 1. The principle of asymmetrical potential wells. (a) Potential energy profiles. (b) Phase profiles. (c) Intensity distribution.

Firstly, we demonstrate controllable asymmetrical potential wells in holographic optical tweezers, which are generated by different phase gradient in ring optical patterns. The holograms are designed by our proposed method, and then the reconstructed phase profiles are obtained, as shown in Fig. 2(a), in which the phase gradient increases from panel I to III. It finds that the phase profiles are very accurate. Additionally, the reconstructed intensity distributions are almost uniform along ring orbit, as given in Fig. 2(b). Consequently, its potential wells are determined by phase gradient profiles. Fig. 2(c) presents the corresponding simulated potential wells, which have different asymmetrical potential gradients. Specially, it can be dynamically controlled by employing different holograms in optical trapping systems. Then, their trapping stiffness are investigated by experimentally monitoring the position fluctuation of trapped nanoparticles. In Fig. 2(d), we present the time trajectories of x- and y-positions of Au nanoparticles with diameter of 200 nm, while trapped by different potential wells. Through numerical analysis, the standard deviations of fluctuation amplitudes are 307 nm, 241 nm and 124 nm in x direction, and are 376 nm, 262 nm and 161nm in y direction, respectively. It indicates that the larger stiffness of potential wells leads to the smaller standard deviation. Accordingly, its trapping confinement can be dynamically controlled by tuning the phase gradient profiles.

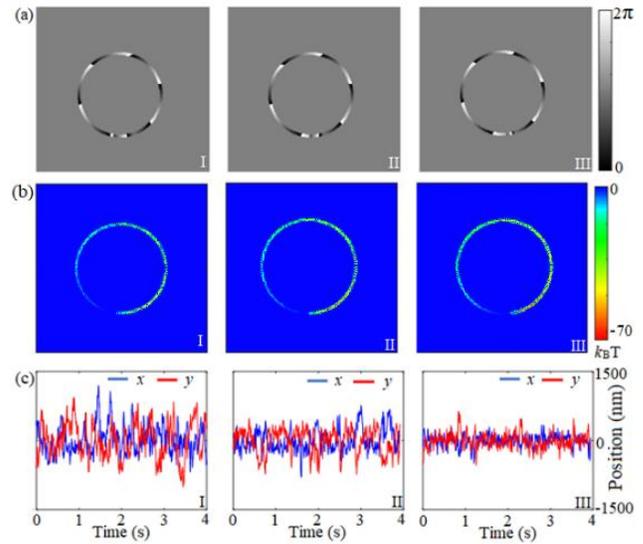

Fig. 2. Controllable asymmetrical potential wells in ring traps. (a) Phase profiles. (b) Potential energy distribution. (c) The positions of the nanoparticles at different times. Note that panels I-III correspond to increasing phase gradients, and the ring radius is about 7 μm.

In the following, we show an new controllable transport by manipulating asymmetrical potential wells, which can synchronously shift nanoparticles from one position to another along circle orbit. Its physical principle is depicted in Figure 3(a). Two different asymmetrical potential profiles, which are marked as I and II, are utilized for manipulating nanoparticles in the different time duration $\Delta T_1$ and $\Delta T_2$. In the beginning, the nanoparticle is assumed to be located at position 1, and then it quickly moves to position 2, where it is trapped. While the potential profile is changed from I to II, its potential energy of nanoparticle is rapidly switched from low (position 2 on profile I) to high (position 3 on profile II). Then, it moves to position 4. In the following, it can be similarly shifted from one location to next, while potential profiles are periodical switched between I and II. The asymmetrical potential wells are given in Figure 3(b), which corresponds to the potential profiles in I and II, respectively. Then, the experiments of controllable transport are carried out. The snapshots of nanoparticle motion at time interval 1.33s are shown in Figure 3(c), and the angular trajectories of three nanoparticles are presented in Figure 3(d). It demonstrates that three nanoparticles can be almost simultaneously shifted from one position to next, all of which have same step distance along circle orbit. Its step distance is about 7 μm, and the average speed is about 21 μm/s. More importantly, its step distance, velocity, path and time interval can be flexibly manipulated by purposely designing desired phase profiles, according to practical requirement. The asymmetrical potential wells has promising applications in optical manipulation such as optical routing, optical sorting and optical Brownian ratchet. Frankly, the symmetrical potential well can be also used for such controllable transport, but at least three holograms are needed for optical manipulation, which could suffer from smaller step distance and slower average speed.

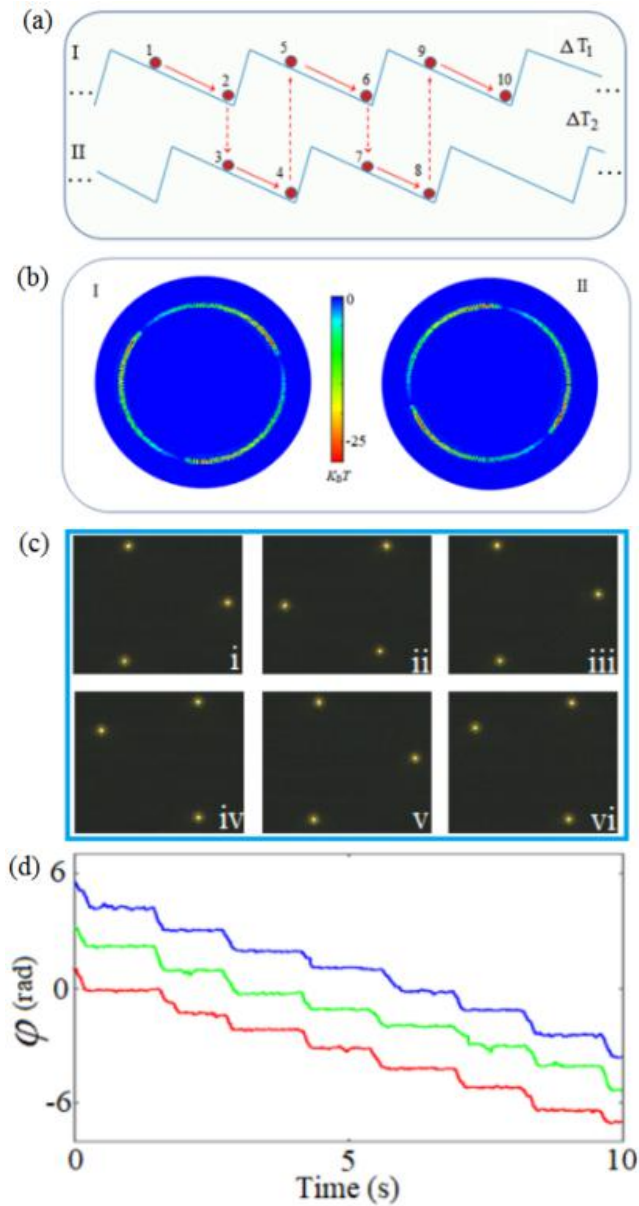

Fig. 3. Controllable transport. (a) Principle of operation. (b) The potential wells at different time. (c) The snapshots of nanoparticles motion at different time. (d) Angular trajectories of three nanoparticles in the ring orbit. Note that panels I-II correspond to different potential distributions, and the ring radius is about 7 μm.

In conclusion, we have proposed a novel approach for variable asymmetrical potential wells, created by utilizing transverse phase gradient profiles. Its depth, stiffness and location of potential wells can be flexibly manipulated, which are highly preferred in optical manipulation. Furthermore, the controllable transport of nanoparticles have been presented and experimentally demonstrated, by employing asymmetrical potential wells, for the first time. The study shows that the manipulation performance is excellent. More importantly, its motion path, step distance and velocity can controlled, by employing the desirable designed holograms. As a result, it opens a new way for developing novel functions in optical manipulation of particles.

**Funding.** Natural Science Foundation of Hunan Province in China (2016JJ2087).

**Disclosures.** The authors declare no conflicts of interest.

**Appendix**
**Experimental Method**
In our experiment system, a CW tunable Ti:Sapphire laser (Spectra-Physics 3900s) with wavelength of 800 nm is operated in a TEM00 Gaussian mode, in which optical polarization is along x direction. A phase-only SLM (Hamamatsu X13138) is employed to generate the optical tweezers using our designed holograms, which belong to 2D optical traps. An inverted microscope (Olympus IX73) with a 60X objective (NA: 1.2, Olympus UPLSAPO 60XW) is utilized for focusing the laser beam and imaging the nanoparticles. A beam profiler (Edmund Optics) is used for capturing the optical intensity of the reconstructed image. The optical power of laser source are about 190 mW in Fig. 1 and 2. The Au nanoparticles (NPs) with diameter of 200 nm are purchased from nanoComposix Inc., which are nearly spherical. The trapped Au NPs are visualized by darkfield microscopy (Olympus U-DCW condenser), and recorded at frame rate of 150 fps, by using a CMOS camera (Point-Grey Grasshopper 3).